\documentclass{emulateapj}

\usepackage{graphicx}
\usepackage{float}
\usepackage{amsmath}
\usepackage{rotating}
\usepackage{subfigure}
\usepackage{epsfig,floatflt}

\begin{document}

\title{Asymmetries in the Cosmic Microwave Background Anisotropy Field} 

\author{H.\ K.\ Eriksen\altaffilmark{1}} \affil{Institute of
Theoretical Astrophysics, University of Oslo, P.O.\ Box 1029 Blindern,
\\ N-0315 Oslo, Norway} \altaffiltext{1}{Also at Centre of Mathematics
for Applications, University of Oslo, P.O.\ Box 1053 Blindern, N-0316
Oslo, Norway}

\email{h.k.k.eriksen@astro.uio.no}

\author{F.\ K.\ Hansen}
\affil{Dipartimento di Fisica, Universit\`a di Roma `Tor Vergata', Via
  della Ricerca Scientifica 1, I-00133 Roma, Italy} 

\email{Frode.Hansen@roma2.infn.it}

\author{A.\ J.\ Banday}
\affil{Max-Planck-Institut f\"ur Astrophysik, Karl-Schwarzschild-Str.\
1, Postfach 1317,\\D-85741 Garching bei M\"unchen, Germany}

\email{banday@MPA-Garching.MPG.DE}

\author{K.\ M.\ G\'orski\altaffilmark{2}} \affil{Jet Propulsion Laboratory, MS
169/327, 4800 Oak Grove Drive, Pasadena CA 91109}
\altaffiltext{2}{Also at Warsaw University Observatory, Aleje
Ujazdowskie 4, 00-478 Warszawa, Poland}

\email{Krzysztof.M.Gorski@jpl.nasa.gov}

\and

\author{P.\ B.\ Lilje\altaffilmark{1}} \affil{Institute of Theoretical
Astrophysics, University of Oslo, P.O.\ Box 1029 Blindern, \\N-0315
Oslo, Norway}

\email{per.lilje@astro.uio.no}

\date{Received 2003 July 30 / Accepted 2003 December 29}

\begin{abstract}
We report on the results from two independent but complementary
statistical analyses of the \emph{WMAP} first-year data, based on the
power spectrum and $N$-point correlation functions.  We focus on large
and intermediate scales (larger than about $3^{\circ}$) and compare
the observed data against Monte Carlo ensembles with \emph{WMAP}-like
properties. In both analyses, we measure the amplitudes of the
large-scale fluctuations on opposing hemispheres and study the ratio
of the two amplitudes. The power-spectrum analysis shows that this
ratio for \emph{WMAP}, as measured along the axis of maximum
asymmetry, is high at the 95\%--99\% level (depending on the
particular multipole range included). The axis of maximum asymmetry of
the \emph{WMAP} data is weakly dependent on the multipole range under
consideration but tends to lie close to the ecliptic axis. In the
$N$-point correlation function analysis we focus on the northern and
southern hemispheres defined in ecliptic coordinates, and we find that
the ratio of the large-scale fluctuation amplitudes is high at the
98\%--99\% level. Furthermore, the results are stable with respect to
choice of Galactic cut and also with respect to frequency band. A
similar asymmetry is found in the \emph{COBE}-DMR map, and the axis of
maximum asymmetry is close to the one found in the \emph{WMAP} data.
\end{abstract}

\section{Introduction}

In recent months much interest has focused on the question of the lack
of large-scale power in the \emph{WMAP}\footnote{Wilkinson Microwave
Anisotropy Probe} (Bennett et al.\ 2003a) first-year data sets (e.g.,
Efstathiou 2003a, b; Efstathiou 2004; Spergel et al.\ 2003; Tegmark, de
Oliveira-Costa, \& Hamilton 2003). Evidence for non-Gaussian features
in the sky maps has also been claimed (Coles et al.\ 2004; Naselsky et
al.\ 2004; Vielva et al.\ 2004; Copi, Huterer, \& Starkman 2004).  In
this paper we present evidence of a related effect, namely that the
large-scale power in \emph{WMAP} is unevenly distributed on the
sky. This effect can be seen on angular scales down to 3--$5^{\circ}$,
or $\ell \lesssim 40$. Similar results based on the genus statistic
have also been reported by Park (2004).

\section{Data and Simulations}

All results in this paper are derived from the \emph{WMAP} first year
Q-, V- and W-band sky maps, corrected for foreground emission using
the template method of Bennett et al.\ (2003b).  The power spectrum
analysis includes the co-added V- and W-band information, while the
$N$-point correlation function analysis utilizes the Q-, V- and W-band
data both separately and in a co-added form. The maps are combined
using inverse-variance noise weights as described by Hinshaw et al.\
(2003b).

The significance of our analyses are evaluated by means of Monte Carlo
simulations: we generate an ensemble of stochastic realizations of the
CMB sky to which the observed data are compared through the power
spectrum and $N$-point statistics.  The CMB simulations are based on
the \emph{WMAP} running-index spectrum. From each CMB realization, we
generate simulated skies corresponding to the eight \emph{WMAP} Q-, V-
and W-band sky maps by convolution with the appropriate beam transfer
functions (Page et al.\ 2003) and then adding a corresponding noise
realization, based on the specific noise properties of the channel in
question.

\section{Analysis Methods}
\label{sec:definitions}

In this paper we analyze the \emph{WMAP} data by means of both the
angular power spectrum (Hansen, G\'orski, \& Hivon 2002) and its
real-space complements, the $N$-point correlation functions (e.g.,
Eriksen, Banday, \& G\'orski 2002). While we do not aim at presenting
the full statistical machinery behind these methods here, instead
referring the interested reader to the above papers for details, we do
provide the specifics appropriate for the \emph{WMAP} assessment.

\subsection{The angular power spectrum}

The angular power spectrum can be determined using a maximum
likelihood inference from a local estimate on a given patch of the
sky, assuming that the coupling between harmonic modes due to the
incomplete sky coverage can be evaluated. Hansen et al.\ (2002) have 
demonstrated that for an axisymmetric patch, this correlation
matrix can be computed analytically. In this paper, we utilize
a Monte Carlo approach to calculate the matrix on a general 
nonsymmetric sky patch. 

Using this likelihood technique, we have estimated the power spectrum 
on 164 slightly overlapping discs with radius $9\fdg5$ uniformly
distributed on the part of the sphere outside of the \emph{WMAP} 
Kp2 sky cut. In particular, we have compared the disc spectra in the
northern Galactic hemisphere to those in the southern sky 
for the multipole bin $\ell=2-63$.

We extend the method further by estimating the power spectra directly
for hemispheres, thus allowing us to obtain a higher multipole
resolution on large scales (low-$\ell$). In order to avoid the
introduction of bias through consideration of hemispheres defined only
with respect to the Galactic coordinate system, we estimated the
spectra for a number of hemispheres defined by orienting the
north-south axis in 82 different directions on the sphere.
Specifically, we applied the following procedure both for the
\emph{WMAP} co-added V- and W-bands, plus each of 2048 simulations.
\begin{itemize}
\item The power spectrum was estimated in bins $C_b$ of width 3, where
  the multipole bin
  $b\in\{\ell^b_{\textrm{min}},\ell^b_{\textrm{max}}\}$ is defined by
  $\ell_{\textrm{min}}=2+3b$ and $\ell_{\textrm{max}}=4+3b$. The spectrum
  was successively estimated on the northern and southern hemispheres
  with the axis defining north and south oriented in 82 different
  directions.
\item The estimated bins $C_b$ in a given multipole range
  $\ell=\ell_\mathrm{min}-\ell_\mathrm{max}$ were summed for each of
  the 164 hemispheres giving $C_\mathrm{hemisphere}=\sum_b C_b$.  The
  ratio between the northern and southern spectra for each of the 82
  possible orientations $r=\textrm{max}(C_N/C_S,C_S/C_N)$ was
  determined.
\item The axis with the maximum asymmetry ratio, $r_\mathrm{max}$,
  was recorded.
\end{itemize}

Subsequently, histograms of the asymmetry ratio were used to quantify
the level of agreement between the simulations and the real sky
measurements.  The results are presented in \S
\ref{sec:results_pwrspec}.

\subsection{$N$-point correlation functions}

An $N$-point correlation function is determined from the average
product of $N$ temperatures, as measured in a fixed relative
orientation defined by an $N$-point polygon on the
sky. Algorithmically speaking, these functions are estimated as simple
averages over all sets of pixels fulfilling the geometric requirements
set by the polygon.

In this paper, the correlation function analyses have been designed
to probe two different ranges of scales.

The very largest scales are analyzed by measuring the three- and
four-point correlation functions from the full sky in $1^{\circ}$
bins, for a total of 120 (equilateral and pseudo-collapsed)\footnote{A
``pseudo-collapsed'' triangle is in this paper defined as an isosceles
triangle with a baseline between 1 and $2^{\circ}$; see Gazta{\~ n}aga
\& Wagg (2003).}  three-point configurations and 235 (rhombic,
2+2-point and 1+3-point; see Eriksen et al.\ 2002) four-point
configurations. To facilitate this analysis, the large-scale $N$-point
correlation functions are computed from low-resolution maps, which are
constructed by first de-convolving the initial \emph{WMAP} beam
transfer function then convolving with a Gaussian beam of
FWHM$=$140\arcmin.  The maps, initially at a
HEALPix\footnote{http://www.eso.org/science/healpix/} resolution of
$N_{\textrm{side}}=512$, are then reconstructed at
$N_{\textrm{side}}=64$, applying suitable corrections for the pixel
window functions at the two resolutions.  A best-fit monopole, dipole
and quadrupole are subtracted from the sky maps before the $N$-point
correlation functions are determined on that part of the sky defined
by the \emph{WMAP} Kp0 mask. However, since our processing smooths the
maps with a relatively wide Gaussian beam, residual, inadequately
corrected foreground structure in the Galactic plane could also be
smoothed beyond the boundaries of the standard Kp0 mask. For this
reason, we extend the mask accordingly in all directions before
proceeding. This extended mask no longer includes the point source
exclusion regions since the smoothing process minimizes their
contribution to our analysis.

The motivation for removing the quadrupole is simply that the observed
quadrupole amplitude is apparently low in the WMAP data, and whether
this is a real feature or simply the outcome of inadequate foreground
removal remains unresolved (see, e.g., Efstathiou 2003b; Efstathiou
2004).  Since correlation functions are intrinsically more sensitive
to low-$\ell$ multipoles and it is difficult to include the
uncertainties related to foreground removal, we choose to eliminate
any concern that this effect could dominate other interesting results.

The two-point correlation function is also evaluated but it has proved
difficult to define an asymmetry parameter in this case, and the
results are therefore omitted in this paper.  Nevertheless, the
four-point functions are compared against the value inferred by the
two-point function of the same sky, as described by Eriksen et al.\
(2002).

\begin{figure}
\center
%\epsscale{0.4}
%\plotone{f2.eps}
%\mbox{\epsfig{file=f1.eps,width=88mm}}
\mbox{\epsfig{file=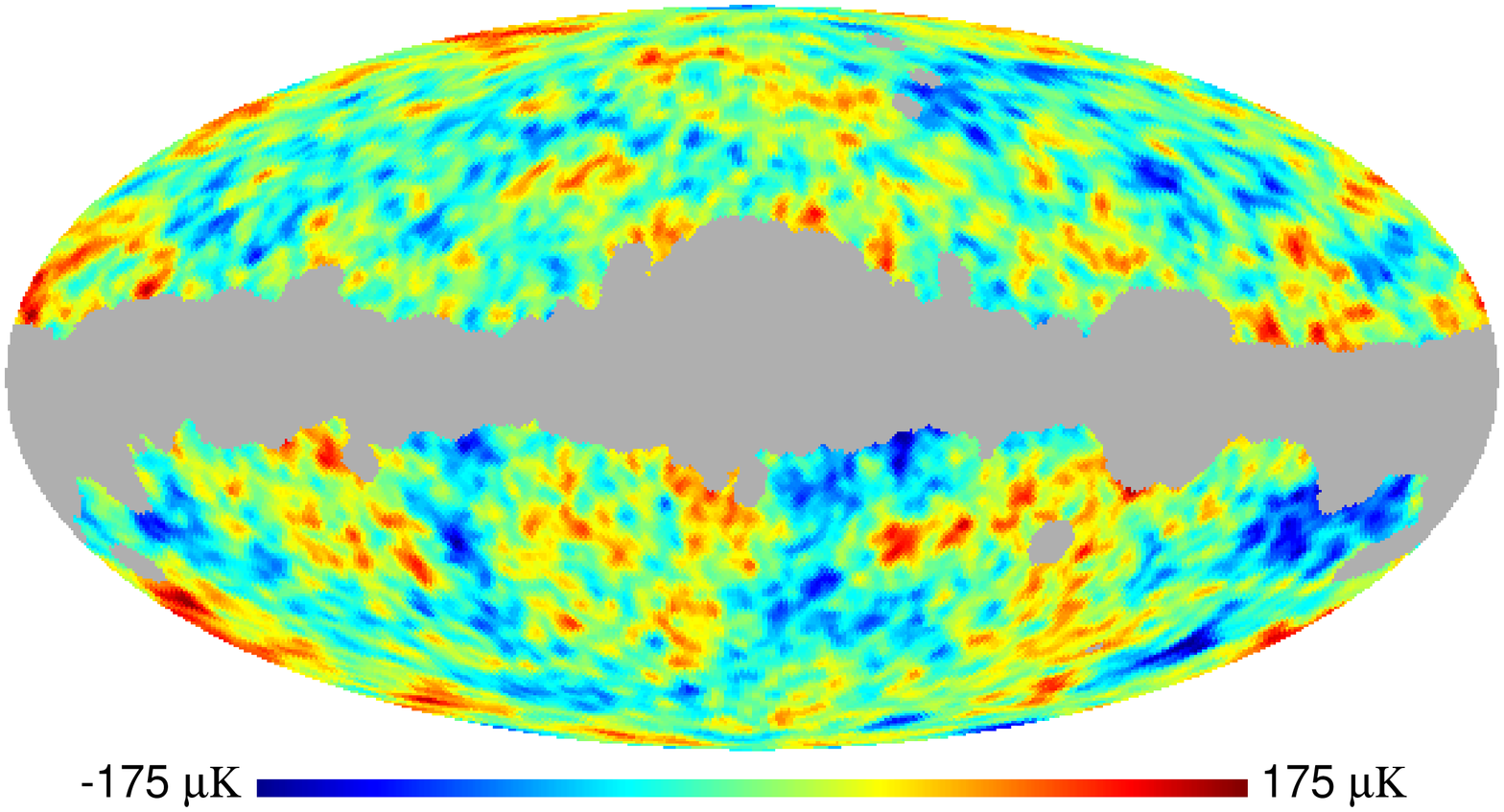,height=88mm,angle=90}}

\vspace*{-4mm}
\mbox{\epsfig{file=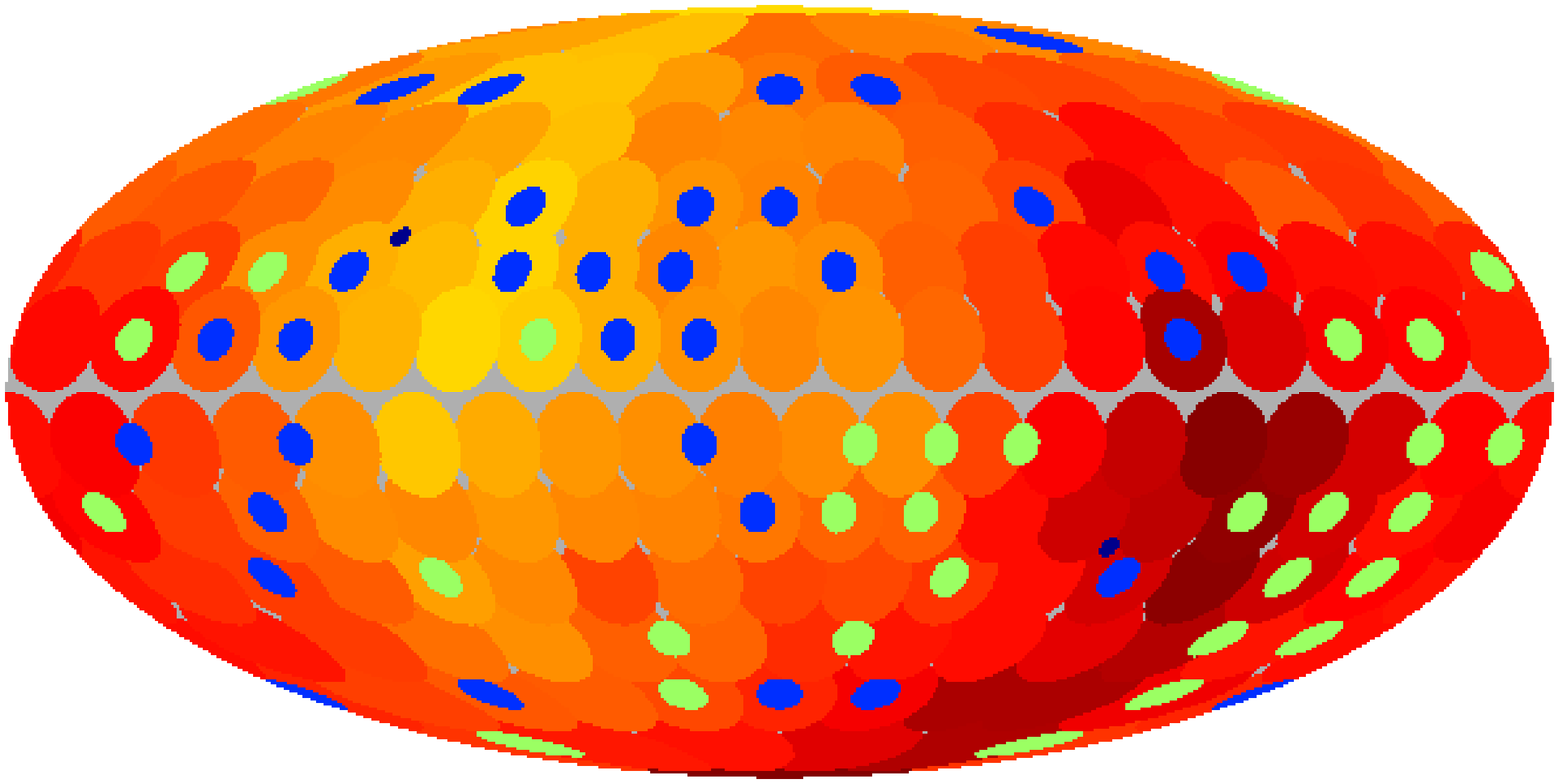,height=88mm,angle=90}}

\vspace*{-6mm}
\mbox{\epsfig{file=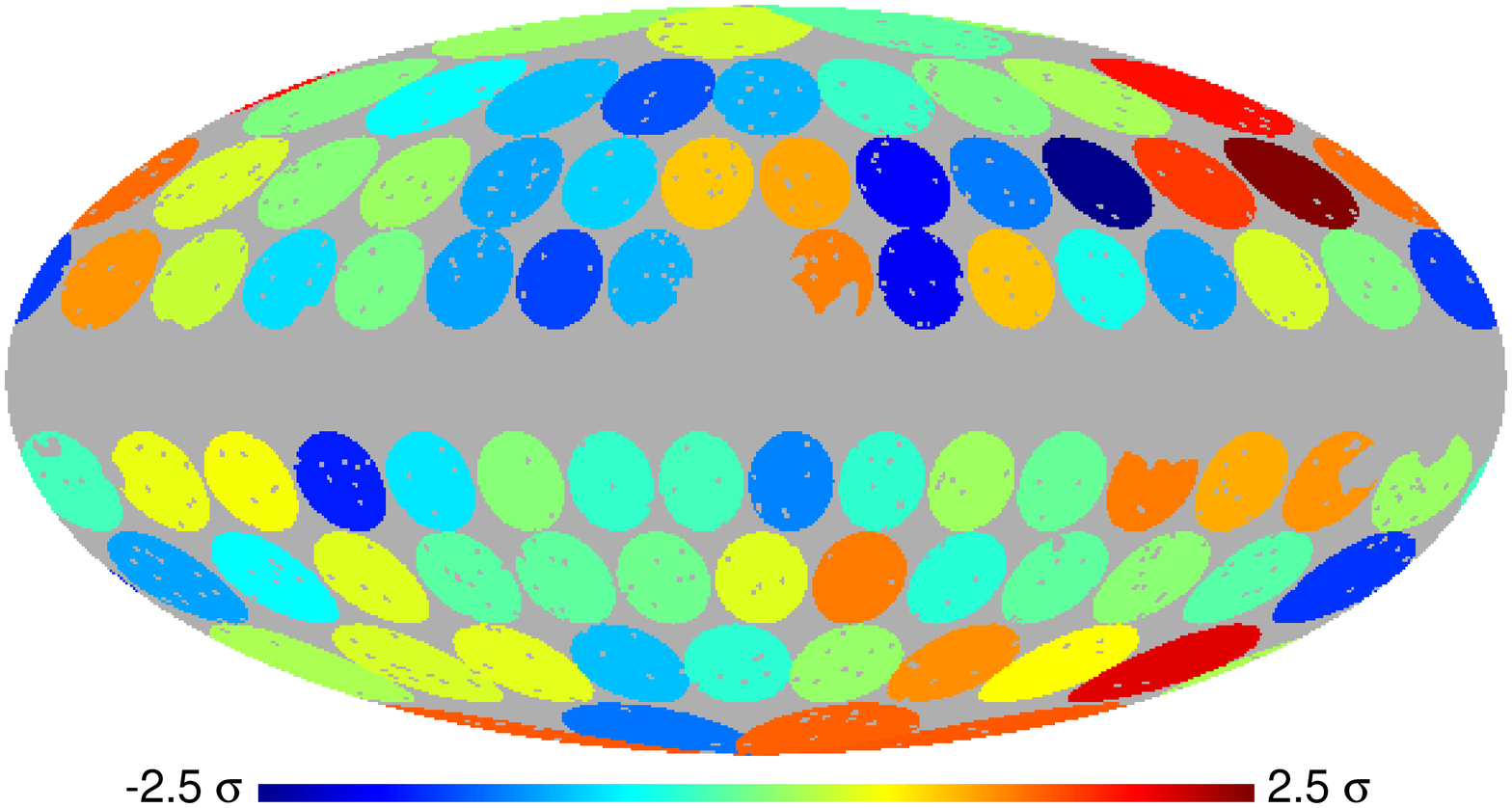,height=88mm,angle=90}}

\caption{
  Top panel: The low-resolution Q-, V- and W-band co-added 
  \emph{WMAP} map, to which the extended Kp0 mask has been applied. \hfill\newline
  Middle panel: Results from the local power 
  spectrum analysis. The color of the large discs indicate the
  ratio between the $\ell=2$--63 power spectrum bin of the northern
  and southern hemispheres, as determined in a reference frame where 
  the north pole pierces the center of the disc;
  light red/yellow indicates a low ratio, dark red a high ratio. 
  The medium sized dots 
  indicate the absolute value of the power spectrum estimated on a $9\fdg5$
  disc in the $\ell=2$--63 bin (i.e., not on the full hemisphere), and
  a dark blue dot means that this value lies below the
  lower $80\%$ confidence limit, while a green dot means it
  lies above the upper $80\%$ limit. Finally, the
  ecliptic poles are marked by the small dark-blue spots. The figure 
  should not be interpreted as implying that data close to
  the galactic plane have been used in the analysis: the \emph{WMAP} Kp2
  mask has been applied for all power spectrum computations. \hfill\newline
  Bottom panel: The results from the intermediate scale three-point
  analysis. Blue corresponds to low $\chi^2$, which again corresponds
  to small fluctuations in the three-point function, while red
  corresponds to high $\chi^2$. The full Kp0 mask has been applied to
  the data.}
\label{fig:wmap}
\end{figure}

Intermediate scales between 3 and $5^{\circ}$ are probed by
partitioning the sphere into 81 discs, each of $10^{\circ}$ radius,
and computing the three-point correlation function for each disc. The
full Kp0 mask is applied to the sky map, including the regions of
exclusion related to known point sources. In order to reduce disc-disc
correlations we first apply a high-pass filter to the maps, removing
all multipoles with $\ell=0,\ldots,18$. On each of these discs, we
compute the three-point correlation functions for 460 isosceles
configurations smaller than $5^{\circ}$.

The degree of agreement between the simulations and the observations
are quantified in terms of a standard covariance matrix $\chi^2$
statistic. Such a statistic is, in principle, only appropriate if the
data under consideration follow a joint Gaussian
distribution. Usually, it also works quite well for mildly
non-Gaussian distributions, and in particular symmetric ones, but for
strongly asymmetric distributions it is likely to yield biased
results. It is easily seen that the distributions for the even-ordered
correlation functions for a given geometrical configuration are in
general strongly asymmetric. We therefore transform the data of each
configuration into a Gaussian distribution by means of the empirical
distribution function, before performing the $\chi^2$ analysis. The
transformation is defined as follows
\begin{equation}
\frac{\textrm{Rank of observed map}}{\textrm{Total number of maps}+1}
= \frac{1}{\sqrt{2\pi}} \int_{-\infty}^{s}
e^{-\frac{1}{2} t^2} dt.
\label{eq:gaussianize2}
\end{equation}
The left-hand side is the fraction of simulations with a lower
correlation function value than the map under consideration (i.e., it
approximates the true, but unknown, cumulative distribution function),
and the right-hand side yields the corresponding value, $s$, measured
relative to a standard normal distribution.

\section{Results}
\label{sec:results}

\subsection{Power spectrum}
\label{sec:results_pwrspec}

We have computed a local power spectrum estimate for 164 slightly
overlapping discs with radius $9\fdg5$, uniformly distributed on the
part of the sphere outside of the \emph{WMAP} Kp2 sky cut and compared
these to spectra derived from an ensemble of 6144 simulated maps.
Concentrating on the lowest multipole bin $\ell=2-63$ we found that
the amplitudes for discs in the northern Galactic hemisphere were
generally lower in the \emph{WMAP} data than in the simulated maps.
Conversely, we found that the discs in the southern Galactic
hemisphere were of generally higher amplitude than in the
simulations. By considering the ratio of the mean of the spectra in
the northern hemisphere to that in the southern hemisphere, we found
that only $0.5\%$ of the simulations have a ratio as low as the
\emph{WMAP} data.  This is the first evidence of a large scale absence
of power in one hemisphere of the \emph{WMAP} data.

\begin{deluxetable}{lcccc}
\tabletypesize{\small}
\tablecaption{Power spectrum asymmetry ratio results\label{tab:bins}}
\tablecomments{
  A summary of results for the ratio of power
  spectrum amplitudes between the northern and southern hemispheres over various $\ell$-ranges   
  defined over the \emph{WMAP} co-added V- and W-band data after applying
  the Kp2 mask. \hfill\newline 
  {\it First column:}  The ratio as computed in the ecliptic
  coordinate frame. The numbers indicate the fraction of simulations
  with a higher asymmetry 
  ratio for the ecliptic axis than the data.   \hfill\newline
  {\it Second column:} 
  The ratio as computed in a coordinate frame selected such that the
  observed asymmetry is maximized for the observational data alone.
  The data values are compared against values from the simulations that have
  the preferred axis imposed on them by the data.
  The numbers indicate the fraction of simulations with a higher asymmetry
  ratio for this axis than the data. \hfill\newline
  {\it Third column:} 
  The ratio as computed in a coordinate frame selected such that the
  observed asymmetry is maximized. 
  The numbers indicate the fraction of simulations with a higher maximum
  ratio $r$ than that found in the \emph{WMAP} or \emph{COBE}-DMR data.
  Note that in this case the data value is compared against values
  derived from the simulations that may or may not have the same
  preferred axis as the data. \hfill\newline
  {\it Fourth column:} The $(\theta,\phi)$ direction of the north pole
  in the galactic reference frame for the axis that maximizes
  the asymmetry observed in the data.
}
\tablewidth{0pt}
\tablecolumns{6}
\tablehead{& $\hat{n}_{\textrm{ecl}}$ & $\hat{n}_{\textrm{max}}$ &
  $P_{\textrm{max}}$  & $(\theta,\phi)_{\textrm{max}}$}
%\tablehead{& ecl. axis & max axis & max ratio & max pos.}
\startdata
WMAP, $\ell=\;\;\,2-40$  & 0.000 & 0.000  & 0.003 & (0,0) \\
WMAP, $\ell=\;\;\,5-40$  & 0.008 & 0.000 & 0.007 & (80,57) \\
WMAP, $\ell=\;\;\,8-40$  & 0.001 & 0.000 & 0.047 & (80,57) \\
WMAP, $\ell=\,20-40$  & 0.000 & 0.000 & 0.009 & (80,57) \\
WMAP, $\ell=\;\;\,2-19$  & 0.048 & 0.000 & 0.002 & (0,0) \\
DMR,  $\;\;\;\ell=\;\;\,2-19$ & 0.037  & 0.001 & 0.131 & (80,95) 
\enddata

\end{deluxetable}

In order to pursue this effect further, we have computed the ratio of
the power spectrum amplitudes determined for the northern and southern
hemispheres as defined in a particular coordinate system, and for a
selection of multipole ranges.  The results are reported in Table
\ref{tab:bins}.  In particular, we consider this ratio after first
determining that coordinate frame that maximizes its value.  The ratio
for the \emph{WMAP} data is larger than $\sim 99\%$ of the maximum
asymmetry values determined from the simulated maps.  We have also
tabulated the orientation, in Galactic coordinates, of the north pole
for the data-preferred reference frame. Note that whereas the
asymmetry on the lowest multipoles seems to be concentrated about the
north Galactic pole, the asymmetry in the higher multipoles
($5<\ell<40$) seems to be highest about the axis with the north pole
at $(\theta,\phi)=(80^\circ,57^\circ)$ in Galactic
coordinates.\footnote{ Here, $\theta$ and $\phi$ are measured in the
HEALPix convention, thus corresponding to \emph{co}-latitude and
longitude.} Such a result may argue against an explanation in terms
of residual foreground contamination, at least for the higher
multipole ranges. It is also clear that the observed asymmetry is not
simply a reflection of the possible low quadrupole and octopole
amplitudes found by the \emph{WMAP} team.  The middle panel of
Figure~\ref{fig:wmap} summarizes these results in a different
way. Each observed disc on the map represents the statistical
deviation, as compared to simulations of the observed asymmetry ratio
when computed in the reference frame for which the north pole
pierces the center of the disc. What is most immediately evident is
that there is an apparent lack of large-scale power in the vicinity of
the north ecliptic pole.

\begin{figure}
%\epsscale{1.0}
%\plotone{f3.eps}
\mbox{\epsfig{file=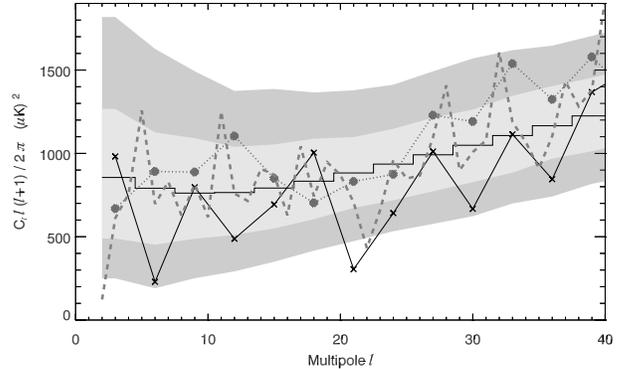,height=88mm,angle=90}}
\caption{Power spectra computed from the co-added V- and W-band
  \emph{WMAP} data.
  The solid line (histogram) indicates the theoretical best-fit
  \emph{WMAP} running index power spectrum.
  The dashed line shows the estimated power spectrum obtained
  by the \emph{WMAP} team for the Kp2 mask. 
  The black crosses and gray dots represent
  our estimates of the power spectra on the northern and southern
  hemispheres, respectively. 
  Here, north and south are defined with respect to the axis that maximizes
  the asymmetry in the \emph{WMAP} data for the corresponding hemispheres, 
  such that the north pole is located at  
  $(\theta,\phi)=(80^\circ,57^\circ)$. 
  The gray bands indicate the 1 and $2\sigma$ confidence regions, 
  as computed from the ensemble of 2048 Monte Carlo simulations.
  Formally, these error bounds differ between the hemispheres, but
  in practice, the difference is small and only the values from the northern
  hemisphere are shown. }
\label{fig:powerspectrum}
\end{figure}

Figure~\ref{fig:powerspectrum} compares the nearly full-sky power
spectrum computed by the \emph{WMAP} team to the local northern and
southern hemisphere estimates derived in the reference frame that
maximizes the asymmetry between them for the multipole range $\ell = 5
- 40$.  This figure also shows the best-fit running-index spectrum.
We see that the northern spectrum is systematically lower than the
southern spectrum over almost the entire multipole range.
				 
As a useful cross check that should help to mitigate against
systematic effects as the cause of the observed structure, the same
hemisphere exercise has been performed for the co-added 53+90 GHz
\emph{COBE}-DMR map, for which we consider multipoles in the range
$\ell=2-19$ where the signal is dominant. We find that the DMR axis of
maximum asymmetry lies close to that for the \emph{WMAP}
data. However, for DMR the significance of the result is lower at
about $87\%$ confidence. Nevertheless, given the noisier nature of the
data, we consider that this is supportive of the asymmetry result.

\subsection{$N$-point correlation functions}

\begin{figure}
%\epsscale{0.8}
%\plotone{f1.eps}
\mbox{\epsfig{file=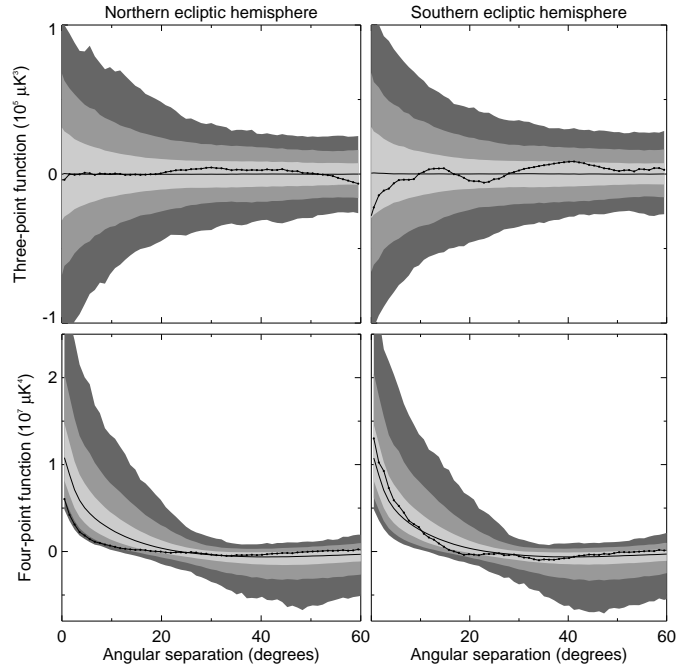,width=88mm}}
\caption{
  Pseudo-collapsed three-point and 1+3 four-point correlation
  functions (dotted line) as computed from the low-resolution co-added Q-, V- and
  W-band \emph{WMAP} data for the Kp0 Galactic mask after imposing an
  additional constraint that $|b|>30^{\circ}$. 
  The gray bands indicate 1, 2, and $3\sigma$
  confidence regions as computed from a set of 5000 Monte Carlo
  simulations. The solid line shows the median.}
\label{fig:corrfuncs_large}
\end{figure}

In Figure \ref{fig:corrfuncs_large}, the pseudo-collapsed three-point
and 1+3 four-point functions are shown as computed for the northern
and southern ecliptic hemispheres. While the expectation values of the
two functions are very different, the observed behavior of the two
functions is internally consistent: the northern hemisphere
correlation functions are strikingly featureless (the three-point
function lies very close to zero, and the four-point function drops
off very quickly), while the southern hemisphere functions show
relatively strong fluctuations.

\begin{deluxetable}{lllcccc}
\tablewidth{0pt}
\tabletypesize{\small}
\tablecaption{$N$-point correlation function $\chi^2$
results\label{tab:chisq_fullsky}}
\tablecomments{Results from $\chi^2$ tests of the large-scale correlation
  functions computed from the northern and southern ecliptic
  hemispheres and from the ratio of northern to southern $\chi^2$ for
  the given bands. Reported here are the frequencies of simulated realizations
  with a $\chi^2$ value ($\chi^2$ ratio) lower (smaller) than for the \emph{WMAP}
  map. The $|b| > 30^{\circ}$ constraint is enforced in addition to
  the Kp0 mask.}
\tablecolumns{6}
\tablehead{ & Mask & Q & V & W & Co-added}
\startdata

\cutinhead{Three-point function}
Northern                 & Kp0 & 0.047 & 0.023 & 0.014 & 0.034 \\
Southern                 & Kp0 & 0.831 & 0.861 & 0.871 & 0.840 \\
Ratio of $\chi^2$'s      & Kp0 & 0.031 & 0.014 & 0.012 & 0.023 \\[3mm]

Northern                 & $|b|>30^{\circ}$ & 0.046 & 0.033 & 0.038 & 0.041 \\ 
Southern                 & $|b|>30^{\circ}$ & 0.792 & 0.820 & 0.829 & 0.793 \\
Ratio of $\chi^2$'s      & $|b|>30^{\circ}$ & 0.040 & 0.031 & 0.031 & 0.039 \\

\cutinhead{Four-point function}
Northern                 & Kp0 & 0.070 & 0.054 & 0.050 & 0.058 \\
Southern                 & Kp0 & 0.852 & 0.873 & 0.846 & 0.857 \\
Ratio of $\chi^2$'s      & Kp0 & 0.030 & 0.021 & 0.025 & 0.025 \\[3mm]

Northern                 & $|b|>30^{\circ}$ & 0.020 & 0.010 & 0.012 & 0.014 \\
Southern                 & $|b|>30^{\circ}$ & 0.927 & 0.942 & 0.931 & 0.932 \\
Ratio of $\chi^2$'s      & $|b|>30^{\circ}$ & 0.004 & 0.001 & 0.002 & 0.002 

\enddata
\end{deluxetable}

In order to quantify these statements, we use the full covariance
matrix $\chi^2$ statistic including all bin-to-bin correlations. The
results from these computations are shown in Table
\ref{tab:chisq_fullsky}.  The first two rows for each mask (Kp0 and
Kp0+$|b|>30\deg$) indicate the frequency of simulations with a {\it
lower} $\chi^2$ value than the \emph{WMAP} data, and the third row
shows the frequency of simulations with a smaller
$\chi^2_{\textrm{north}}/\chi^2_{\textrm{south}}$.  This latter
statistic merits some explanation. The $\chi^2$ statistic in itself
measures the overall consistency of an observed function with a
predefined model relative to the standard deviation of the model.
Thus, for a function with vanishing mean (such as the three-point and
the \emph{reduced} four-point functions) the $\chi^2$ statistic is a
monotonic function of the overall fluctuation level. Therefore, we
choose the $\chi^2$ itself as a measure of the fluctuation amplitudes,
and quantify the degree of asymmetry between the two hemispheres by a
parameter that is then the ratio of the two $\chi^2$ values.

The conclusions to be drawn from Table \ref{tab:chisq_fullsky} are as
follows. The northern ecliptic hemisphere shows a consistently low
level of fluctuations, independent of frequency and Galactic cut, but
inside the allowed range set by the model sky. The southern hemisphere
shows a slightly high level of fluctuations but is again acceptable
given the model. However, the two are only marginally consistent
internally: as measured by the three-point function only $\sim 2\%$ of
the simulations have a smaller ratio of $\chi^2$ with the Kp0 mask,
and $\sim 4\%$ with the additional constraint that
$|b|>30^{\circ}$.  Somewhat stronger numbers are found by the
four-point function. Again, about $2\%$ of the simulations have a
smaller ratio of $\chi^2$ in the Kp0 region, but this time only
0.1\%--0.4\% have a smaller ratio in the extended region. The latter
values tend to refute the possibility that residual foregrounds could
be responsible for the observed asymmetry. Note also that the
strongest results are found in the bands least affected by
foregrounds, V and W.

The top panel of Figure \ref{fig:wmap} shows the smoothed, co-added
\emph{WMAP} map, and in this figure it is actually possible to see the
asymmetry by eye; while there are conspicuous large-scale structures
visible on the southern hemisphere, the northern hemisphere is
surprisingly featureless. Therefore, in order to localize these
features further, we have computed the three-point correlation
function in 460 isosceles configurations smaller than $5^{\circ}$ on a
set of 81 discs, after subtraction of all power on scales $\ell <
18$. The result is shown in the lower panel of Figure \ref{fig:wmap},
to aid comparison with the power spectrum results shown in the middle
panel.  An asymmetry is also clearly observed on these scales: large
connected regions on the northern hemisphere have a low $\chi^2$,
whereas the southern hemisphere regions (especially in the eastern
quadrant) demonstrate higher values.  The possible alignment of
structure with the ecliptic poles is in very good agreement with both
the $N$-point results derived from the ecliptic hemispheres and the
power spectrum analysis over the range $\ell = 20-40$.

\section{Conclusions}
\label{sec:conclusions}

In this paper we have estimated power spectra and $N$-point
correlation functions from the \emph{WMAP} first-year Q-, V- and
W-band data, and compared these against Monte Carlo ensembles based on
Gaussian realizations of the \emph{WMAP} best-fit running-index power
spectrum. Subsequently, we have found evidence for an anisotropic
distribution of large-scale power, as implied by the fact that the
ratio between the northern and southern ecliptic hemisphere power
spectra is low at the $3\sigma$ level.  When the \emph{WMAP} data are
observed in a reference frame that maximizes the asymmetry between
hemispheres, the observed ratio determined from the power spectrum
over the range $\ell = 2 - 40$ is higher than that computed for at
least 99.7\% of the simulations. This is the case even when the
asymmetry ratio for each simulation is computed in its own specific
reference frame of maximum asymmetry.  The significance of the result
is therefore independent of the reference-frame.  The asymmetry was
also confirmed by an $N$-point correlation function analysis, in
which the ratio of $\chi^2$ values computed from the ecliptic
hemisphere correlation functions is low, typically at the $98\%$ level,
and in some cases, even as low as 0.1\%--0.4\%.

The origin of the asymmetry is most simply expressed as a remarkable
absence of large-scale power in the vicinity of the north ecliptic
pole. Nevertheless, it is also a possibility that fundamentally
non-Gaussian signals contribute. Indeed, the three-point correlation
function in the northern ecliptic hemisphere exhibits what could be
described as ``super-Gaussianity.'' Such a property had previously been
determined for the \emph{COBE}-DMR data with a bispectrum statistic
(Magueijo 2000).  Whether this is the case for \emph{WMAP} remains to
be seen, but such an unexpected lack of deviation from the zero
expectation could be interpreted either way.

It is certain that one criticism of this work could be that, given $N$
statistical tests as applied to a specific data set, some may prove to
be anomalous simply by chance.  The specious nature of this argument
should be clear, since the assertion only applies for tests concerning
the same non-Gaussian degree of freedom, whereas we have presented (at
least) three.  Moreover, it is certainly rendered moot by the
increasing number of recent supporting results in the literature. In
particular, Park (2004) has independently discovered the north-south
asymmetry through the use of genus statistics.  In addition, the low
quadrupole and octopole amplitudes already noted in the
\emph{COBE}-DMR data have been confirmed by the \emph{WMAP} team
itself, but more interestingly, de Oliveira-Costa et al.\ (2004) have
noted a conspicuous alignment of these features.  Finally, several
groups have made claims for the detection of non-Gaussianity with a
variety of tests based on phase (Naselsky et al. 2004; Coles et al.\
2004), wavelets (Vielva et al.\ 2004), and multipole vectors (Copi et
al.\ 2004).

That the \emph{WMAP} data does indeed contain unusual features on
large scales seems inescapable. The main issue now is to elucidate
possible explanations; three possible candidates are that these
effects are due to systematics, foregrounds and exotic physics.

It would appear that the \emph{WMAP} data are remarkably free of
systematic artifacts (Hinshaw et al.\ 2003a).  Moreover, the results
presented in this paper seem unlikely to be compromised by systematics
since we find supporting evidence from similar features in the
\emph{COBE}-DMR maps, which are susceptible to \emph{different}
parasitic signals.  Nevertheless, either previously unknown signals
could be postulated, or one might accept the possibility that the
projection of the known systematic signals onto these new statistical
tests is more complex than expected (cf.\ the \emph{COBE}-DMR
bispectrum situation as discussed by Banday, Zaroubi, \& G\'orski
2000).

Residual foreground contamination also seems unlikely. In particular,
we note that the $\ell=20-40$ power spectrum results find 
an axis of maximum asymmetry such that the north pole
of the coordinate system lies away from the Galactic plane
and the axis separating the hemispheres is highly
inclined to it.
More importantly, the observed asymmetry is remarkably stable
with respect to frequency and sky coverage.

The most intriguing possibility remains therefore that the recent findings
require some fundamentally new physics on large scales in the universe.

\begin{acknowledgements}
FKH acknowledges financial support from the CMBNET Research Training
Network and HKE and PBL acknowledge financial support from the
Research Council of Norway.  Some of the results in this paper have
been derived using the HEALPix (G\'orski et al.\ 1999) software and
analysis package.  We acknowledge use of the Legacy Archive for
Microwave Background Data Analysis (LAMBDA). Support for LAMBDA is
provided by the NASA Office of Space Science.  This research used
resources of the National Energy Research Scientific Computing Center,
which is supported by the Office of Science of the U.S. Department of
Energy under Contract No.\ DE-AC03-76SF00098. This work has also
received support from The Research Council of Norway (Programme for
Supercomputing) through a grant of computing time.
\end{acknowledgements}

\end{document}